\documentclass{elsart}
\usepackage{amssymb}
\usepackage{latexsym}
\usepackage{graphics}
\usepackage{graphicx}
\begin{document}
\begin{frontmatter}
\title{Bayes-optimal inverse halftoning and statistical 
mechanics of the Q-Ising model}
\author[label1]{Yohei Saika}
\ead{saika@wakayama-nct.ac.jp}
\address[label1]{Department of Electrical and Computer Engineering,  
Wakayama National College of Technology, 
Nada-cho, Noshima 77, Gobo-shi, Wakayama 644-0023, Japan}

\author[label2]{Jun-ichi Inoue}
\address[label2]{
Complex Systems Engineering, 
Graduate School of Information Science and Technology, 
Hokkaido University, 
N14-W9, Kita-ku, Sapporo 060-0814, Japan}

\author[label3]{Hiroyuki Tanaka and Masato Okada}
\address[label3]{Division of Transdisciplinary Science, 
Graduate School of Frontier Science, 
The University of Tokyo, 5-1-5 Kashiwanoha, Kashiwa-shi, 
Chiba 277-8561, Japan}
\begin{abstract}
On the basis of statistical mechanics of 
the Q-Ising model, we formulate the Bayesian inference 
to the problem of inverse halftoning, 
which is the inverse process of 
representing gray-scales in images by means of 
black and white dots. 
Using Monte Carlo simulations, 
we investigate statistical properties of 
the inverse process, especially, 
we reveal the condition of 
the Bayes-optimal solution for which 
the mean-square error takes its minimum. 
The numerical result is qualitatively confirmed 
by analysis of the infinite-range model. 
As demonstrations of our approach, 
we apply the method to retrieve a grayscale image, such as standard image 
{\it Lenna}, from the halftoned version. 
We find that the Bayes-optimal solution 
gives a fine restored grayscale image 
which is very close to the original. 
\end{abstract}
\end{frontmatter}
\section{Introduction}
\label{sec:Intro}
In recent two or three decades, 
a considerable number of researchers 
have investigated various problems in information sciences, 
such as image restoration and error-correcting codes 
on the basis of the analogy between statistical mechanics 
and probabilistic information processing \cite{Nishi}. 
Especially, a lot of researchers have 
investigated various problems in image processing 
based on the Markov random 
fields \cite{Bes,Gon,Pry,Winkler}. 
In the field of the print technologies, 
many techniques of information processing 
have also developed. 
Particularly, the {\it digital halftoning} 
\cite{Uli,Bay,Flo,Mic,Won} 
is regarded as a key processing
to convert a digital grayscale image to 
black and white dots which represents the original 
grayscale levels appropriately.  
On the other hand, the inverse process of 
the digital halftoning 
is referred to as 
{\it inverse halftoning}. 
The inverse halftoning is also important for us to 
make scanner machines to retrieve the original grayscale image 
by making use of much less informative materials, such as the halftoned 
binary dots. 
The inverse halftoning 
is `ill-posed' in the sense that 
one lacks information to 
retrieve the original image because 
the material one can utilize is just only 
the halftoned black and white binary dots 
instead of the grayscale one. 
To overcome this difficulty, 
we usually introduce 
the `regularization term' 
which compensates the lack of 
the information and regard 
the inverse problem as a combinatorial 
optimization \cite{Cabrera,Discepoli}. 
Then, the optimization 
is achieved to find the lowest energy state 
via, for example, simulated annealing \cite{Kir,Gem}. 

Besides the standard regularization 
theory, we can use the Bayesian 
approach. Under the direction of this 
approach, 
Stevenson \cite{Ste} attempted 
to apply the maximum of a Posteriori (MAP for short) 
estimation to the problem of inverse halftoning 
for a given halftone binary dots obtained 
by the threshold mask and the so-called error diffusion methods. 
However, there is few theoretical approach 
to deal with the inverse-halftoning 
from the view point of the Bayesian inference and 
statistical mechanics of information. 

In this study, on the basis of statistical mechanics of 
the Q-Ising model \cite{Bolle}, we formulate the problem 
of inverse halftoning to estimate 
the original grayscale levels by using 
the information about both the halftoned binary dots and 
the threshold mask.
We reconstruct the original grayscale revels 
from a given halftoned binary image and 
the threshold mask 
so as to maximize the posterior marginal probability. 
Using Monte Carlo simulations, 
we investigate statistical properties of 
the inverse process, especially, 
we reveal the condition of 
the Bayes-optimal solution for which 
the mean-square error takes its minimum. 
The result of the simulation is supported 
by the analysis of the infinite-range model. 
In order to investigate 
to what extent the Bayesian approach 
is effective for realistic images, 
we apply the method to retrieve the 
grayscale levels of the 256-levels standard image 
{\it Lenna} from the binary dots. 
We find that the Bayes-optimal solution 
gives a fine restored grayscale image 
which is very close to the original one. 

The contents of this paper are organized as follows. 
In the next section, 
we formulate the problem of inverse halftoning. 
We mention the relationship between 
statistical mechanics of the Q-Ising model and Bayesian inference 
of the inverse halftoning. 
In the following section, 
we investigate statistical properties of the Bayesian 
inverse halftoning by Monte Carlo simulations. 
Analysis of the infinite-range model 
supports the result of the simulations. 
We also show that the Bayes-optimal inverse halftoning 
is useful even for realistic images, such as the 256-level 
standard image {\it Lenna}. 
Last section is summary. 
\section{The model}
\label{sec:model}
\begin{figure}[ht]
\begin{center}
\mbox{}\vspace{-0.3cm}
\includegraphics[width=5.0cm]{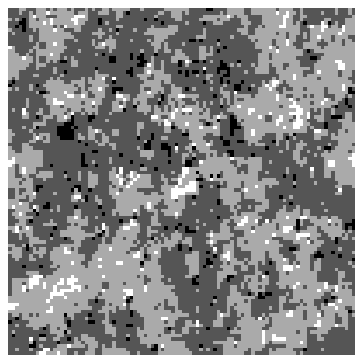} \hspace{1cm}
\includegraphics[width=5.0cm]{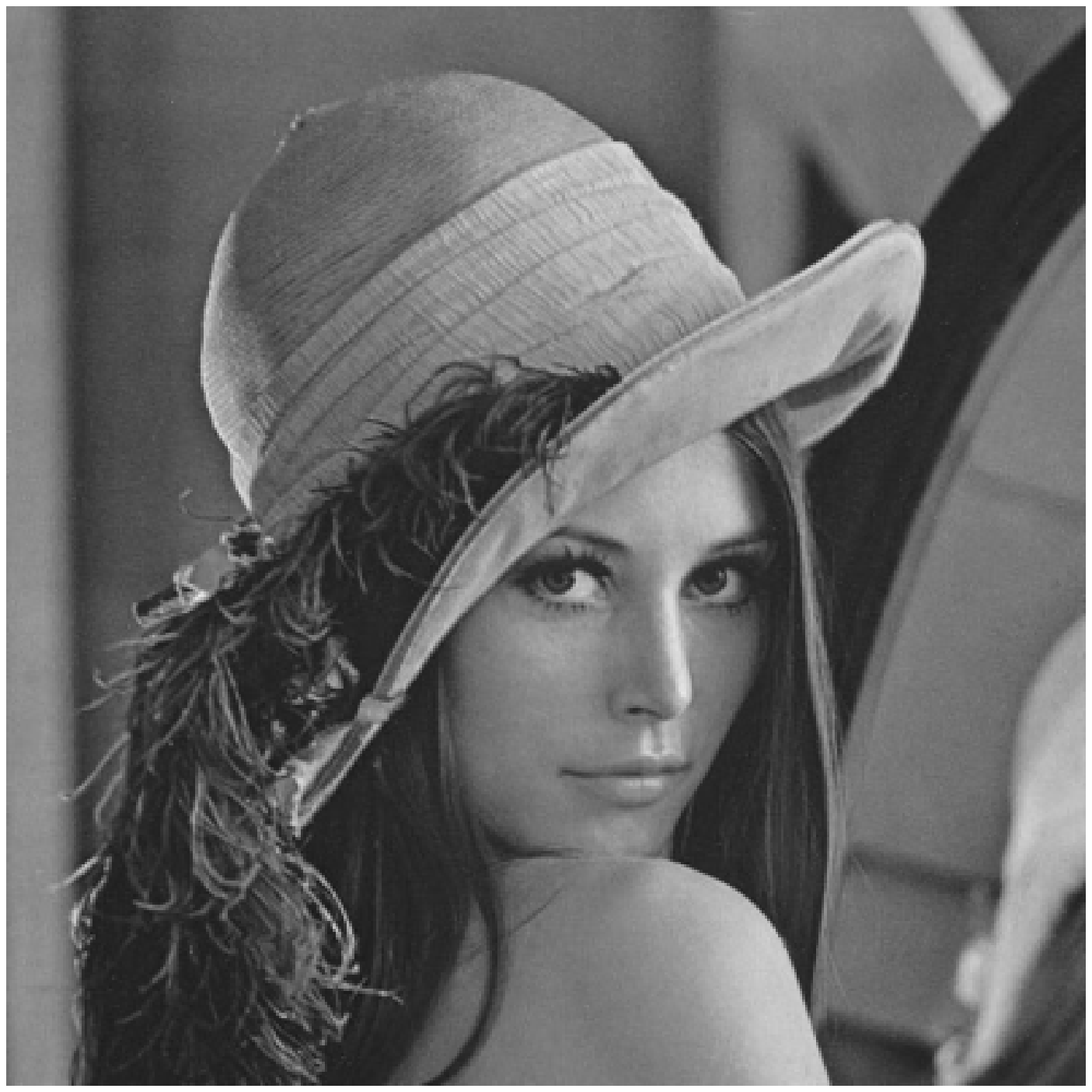}
\mbox{}\vspace{0.5cm}
\end{center}
\caption{\footnotesize 
An original image as a snapshot from 
the Gibbs distribution of (\ref{eq:ferro}) having 
$100 \times 100$ pixels 
for the case of $Q=4$ (left). 
We set $T_s=0.5$, $J=1$. 
The right panel shows 
a $256$-levels standard image {\it Lenna} with $400 \times 400$ pixels. 
}
\label{fig:fg1}
\end{figure}

We first define the model system 
to investigate the statistical performance of 
the Bayesian 
inference for the problem of 
inverse halftoning. 
As original 
grayscale images, which are converted to the black and 
white binary dots, 
we consider snapshots from a Gibbs distribution of 
the ferromagnetic Q-Ising model 
having the spin variables 
$\{ \xi \} \equiv \{ \xi_{x,y} = 0, \cdots, Q-1|x,y=0,\cdots,L-1\}$. 
Then, each image $\{\xi\}$ being 
specified by the Hamiltonian 
$H(\{\xi\})=J_{s} \sum_{n.n.}(\xi_{x,y}-\xi_{x^{'},y^{'}})^{2}$ 
follows the Gibbs distribution 
\begin{equation}
{\rm Pr} \left( \{ \xi \} \right) = 
\frac{1}{Z_{s}}
\exp 
\left[
-\frac{H(\{\xi\})}{T_{s}}
\right]=
\frac{1}{Z_s} 
\exp \left[
- \frac{J_s}{T_s} \sum_{\rm n.n.} 
\left( \xi_{x,y} - \xi_{x^{'},y^{'}} \right)^2 
\right]
\label{eq:ferro}
\end{equation}
at temperature $T_{s}$, 
where $Z_s$ is the partition function 
of the system and 
the summation 
$\sum_{\rm n.n.}(\cdots)$ 
runs over the sets of the nearest neighboring pixels 
located on the square lattice in 
two dimension. 
The ratio of strength of 
spin-pair interaction $J_s$ and 
temperature $T_s$, 
namely, $J_{s}/T_{s}$ controls the smoothness of 
our original image $\{\xi\}$. 
In Fig. \ref{fig:fg1} (left), 
we show a typical example of the snapshots from 
the distribution (\ref{eq:ferro}) 
for the case of $Q=4$, $J_{s}=1$ and $T_{s}=0.5$. 
The right panel of the 
Fig. \ref{fig:fg1} shows the 256-levels grayscale 
standard image {\it Lenna} 
with $400 \times 400$ pixels. 
We shall use the standard image 
to check the efficiency of our approach 
in the last part of this paper. 

In order to convert original grayscale images to the 
the black and white binary dots, 
we make use of the threshold array 
$\{ M \}$. 
\begin{figure}[ht]
\begin{center}
{\large 
\begin{tabular}{|c|c|}
\hline
0 & 2 \\
\hline 
3 & 1 \\
\hline
\end{tabular} 
\mbox{}\hspace{2cm}
\begin{tabular}{|c|c|c|c|}
\hline
0 & 8 & 2 & 10 \\
\hline 
12 & 4 & 14 & 6 \\
\hline
3 & 11 & 1 & 9 \\
\hline
15 & 7 & 13 & 5 \\
\hline
\end{tabular}} 
\end{center}
\caption{\footnotesize  
The Bayer-type threshold arrays 
for the dither method with $2 \times 2$ (left) and with 
$4 \times 4$ (right). 
}
\label{fig:fg2}
\end{figure}
Each component $M_{k,l}$ 
of the array $\{M \}$ takes a non-overlapping integer and 
these numbers are arranged 
on the $L_m \times L_m$ squares as 
shown in Fig. \ref{fig:fg2} for 
$L_m=2$ (left) and for $L_{m}=4$ (right). 
For general case of $L_m$, we define the array as  
\begin{eqnarray}
\{M \} & = & \left\{
M_{k,l}=0,
\frac{Q-1}{L_m^{2}-1},
\frac{2(Q-1)}{L_m^{2}-1}, 
\cdots, Q-1 {\biggr |}
k, l = 0,1,\cdots, L_m-1 
\right\}. \nonumber \\
\label{eq:mask}
\end{eqnarray}
We should keep in mind that 
the definition (\ref{eq:mask}) 
is reduced to $\{M \} =\{
M_{k,l}=0,1,\cdots,Q-1|
k, l = 0,1,\cdots, \sqrt{Q}-1 
\}$
and the domain of each component of 
the threshold array 
becomes the same as that of the original image 
$\{\xi\}$ for $L_{m}^{2}=Q$. 

\begin{figure}[ht]
\begin{center}
\mbox{}\vspace{-0.3cm}
\includegraphics[width=4.5cm]{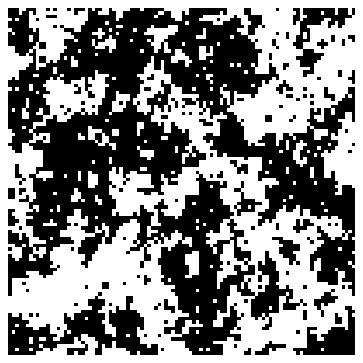}
\includegraphics[width=4.5cm]{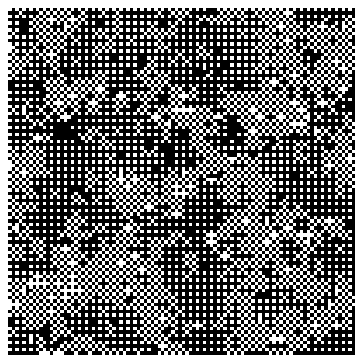}
\includegraphics[width=4.5cm]{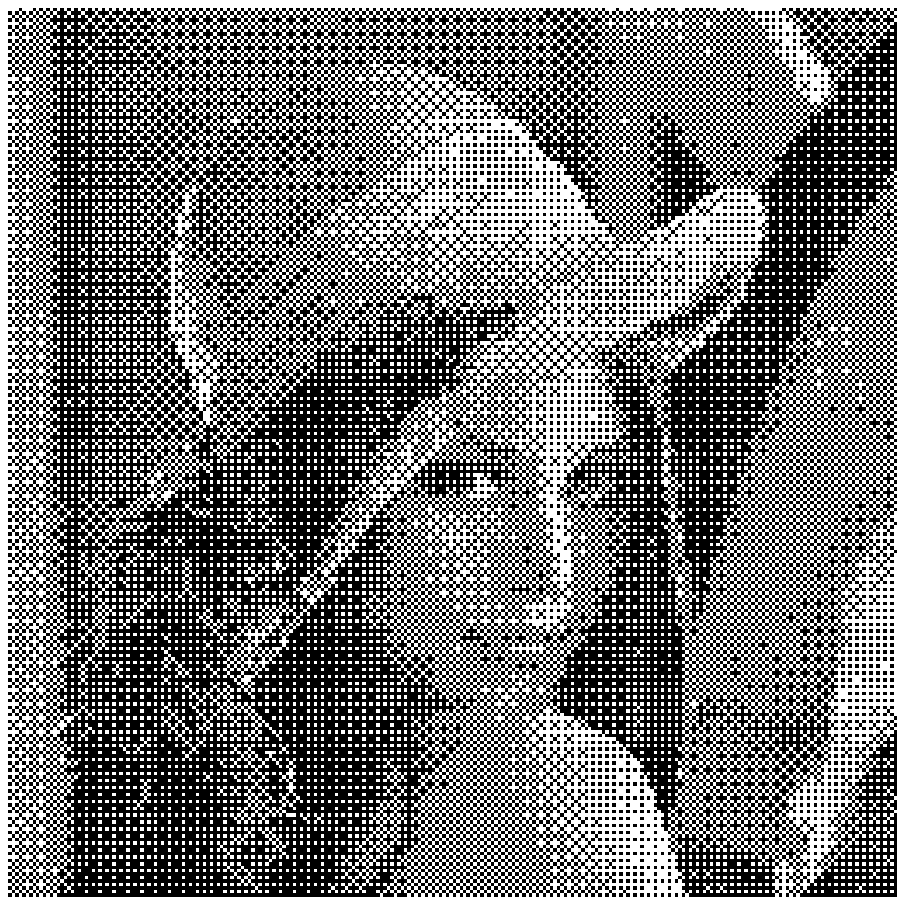}
\mbox{}\vspace{-0.1cm}
\end{center}
\caption{\footnotesize 
The left panel shows 
a halftone image converted by the dither method using the uniform 
threshold $M=2$ from the snapshot 
from a Gibbs distribution 
of the $Q=4$ Ising model 
shown in Fig. \ref{fig:fg1} (left). 
The center panel shows 
a halftone image obtained by the dither method using the $2\times 2$ 
Bayer-type threshold array from the same 
snapshot. 
The right panel shows 
a halftone image converted by the dither method using the $4\times 4$ 
Bayer-type threshold array from the 256-level standard image {\it Lenna} 
with $400 \times 400$ pixels shown in Fig. \ref{fig:fg1} (right). 
}
\label{fig:fg3}
\end{figure}

In order to achieve a pixel-to-pixel 
map between each element of the threshold array, 
$M_{x,y}$ and 
the corresponding original grayscale pixel $\xi_{x,y}$, 
we spread a lots of threshold arrays over the original image 
so as not to overlap any threshold array with one another. 
Then, we transform each original pixel 
$\xi_{x,y}$ into the binary dot $\tau_{x,y}$ by 
\begin{equation}
\tau_{x,y} = 
\theta \left( \xi_{x,y} - M_{x,y} \right)
\label{eq:step}. 
\end{equation}
Here we defined $M_{x,y}$ as the threshold value 
corresponding to the $(x,y)$-th pixel 
and $\theta(\cdots)$ denotes the unit-step function. 
Halftone images generated by the dither method 
via (\ref{eq:step}) 
are shown in Fig. \ref{fig:fg3}. 
We find that the left panel 
obtained by the uniform 
threshold mask $M_{x,y}=2\,\,
(\forall_{x,y})$ is hard to be recognized as a 
grayscale image, whereas, 
the center panel 
obtained by the 
$2 \times 2$ Bayer-type threshold array 
might be recognized as just like 
an original image 
through our human vision systems 
(due to a kind of {\it optical illusion}). 

Obviously, the inverse process of 
the above halftoning is regarded as 
an ill-posed problem. 
This is because from (\ref{eq:step}), 
one can not determine the original 
image $\xi_{x,y}\,(\forall_{x,y})$ 
completely from a given set of 
$\tau_{x,y}\,(\forall_{x,y})$ and 
$M_{x,y}\, (\forall_{x,y})$. 
Then, the standard regularization 
theory \cite{Cabrera,Discepoli} 
provides us a realistic break-through. 
In the theory, we introduce 
the so-called `regularization term' 
that compensates the lack of 
the information to retrieve 
the original image. Then, 
we construct the energy function 
to be minimized to find the original 
image as the lowest energy state. 
For instance, Some recent progress 
based on the standard regularization 
theory is found in our paper \cite{SaikaInoue2008}. 

The standard regularization 
theory is itself a general and powerful 
approach, nevertheless, 
we here use an alternative, 
namely, the Bayesian 
approach to 
solve the inverse problem. 
\mbox{} 

In the Bayesian inverse digital halftoning, 
we attempt to restore the original grayscale image from 
a given halftone image by means of the so-called 
maximizer of posterior marginal (MPM for short) estimate. 
Then, we define 
$\{z\}=\{
z_{x,y} = 0, \cdots, Q-1|x,y=0, \cdots, L-1\}$   
as an estimate of the original image 
$\{\xi\}$ arranged on the square lattice and 
reconstruct the grayscale image on the bases of 
maximizing the following posterior marginal 
probability: 
\begin{eqnarray}
\hat{z}_{x,y} & = & 
\arg \max_{z_{x,y}} 
\sum_{\{ z \} \not= z_{x,y}} 
{\rm Pr} 
( \{ z \} | \{ \tau \}) = 
\arg \max_{z_{x,y}} {\rm Pr} \left( z_{x,y} | \{ \tau \} \right), 
\end{eqnarray}
where the summation 
$\sum_{z_{x,y} \neq \{z\}}(\cdots)$ 
runs over all pixels except for the $(x,y)$-th and 
the posterior probability $P(\{ z \}|\{ \tau \})$ 
is given by the Bayes formula: 
\begin{equation}
{\rm Pr} \left( \{ z \} | \{ \tau \} \right) = 
\frac{
{\rm Pr} \left( \{ z \} \right) {\rm Pr} \left( \{ \tau \} | \{ z \} \right) 
}{
\sum_{\{ z \}}
{\rm Pr} \left( \{ z \} \right) {\rm Pr} \left( \{ \tau \} | \{ z \} \right) 
}
\end{equation}
In this study, following Stevenson \cite{Ste}, 
we assume that 
the likelihood might have the same form as the halftone 
process of the dither method, namely, 
\begin{equation}
P \left( \{ \tau \} \vert \{ \xi \} \right) = 
\Pi_{(x,y)} 
\delta \left( \tau_{x,y}, 
\theta \left( z_{x,y} - M_{x,y} \right) 
\right), 
\end{equation}
where $\delta(a,b)$ denotes 
a Kronecker delta and we should notice that 
the information on 
the threshold array $\{ M \}$ 
is available in addition to the halftone image 
$\{\tau\}$. 
Then, we choose the model of the true prior as 
\begin{equation}
{\rm Pr} (\{ z \}) =  
\frac{1}{Z_{m}} \exp 
\left[
- \frac{ J }{ T_{m} } 
\sum_{\rm n.n.} \left( z_{x,y} - z_{x^{'},y^{'}} \right)^2 
\right], 
\end{equation}
where $Z_{\rm m}$ is a normalization factor. 
$J$ and $T$ are the so-called hyper-parameters. 
It should be noted that 
one can construct the Bayes-optimal solution 
if we assume that the model prior has the same form 
as the true prior, namely, 
$J = J_{s}$ and $T_{m} = T_{s}$ 
(what we call, {\it Nishimori line} 
in the research field of spin glasses \cite{Nishi}). 
\par
From the viewpoint of statistical mechanics, 
the posterior probability 
${\rm Pr}(\{z\}|\{\tau\})$ 
generates the equilibrium states 
of the ferromagnetic Q-Ising model whose Hamiltonian is given by  
\begin{equation}
H \left( \{ z \} \right) = 
J \sum_{\rm n.n.} \left( z_{x,y} - z_{x',y'} \right)^2
\label{eq:Ham}, 
\end{equation}
under the constraints 
\begin{equation}
\forall_{x,y}\,\,\,\,\,\,\,\,
\tau_{x,y} = \theta \left( z_{x,y} - M_{x,y} \right). 
\label{eq:constraint}
\end{equation}
Obviously, 
the number of possible spin configurations 
that satisfy the above constraints (\ref{eq:constraint}) 
is evaluated as $\prod_{(x,y)}|Q\tau_{x,y}
-M_{x,y}|$ and this quantity is exponential order 
such as $\sim {\alpha}^{L^{2}}$ 
($\alpha$: a positive constant). 
Therefore, 
the solution $\{z\}$ to satisfy the constraints (\ref{eq:constraint})
is not unique and this fact makes the problem very hard. 
To reduce the difficulties, 
we consider the equilibrium state generated by a 
Gibbs distribution of the ferromagnetic Q-Ising model 
with the constraints (\ref{eq:constraint}) 
and increase the parameter $J$ gradually from $J=0$. 
Then, we naturally expect that  
the system stabilizes the ferromagnetic 
Q-Ising configurations due to a kind of 
the regularization term (\ref{eq:Ham}). 
Thus, 
we might choose the best possible 
solution among a lot of 
candidates satisfying (\ref{eq:constraint}). 

>From the view point of statistical mechanics, 
the MPM estimate is rewritten by 
\begin{eqnarray}
\hat{z}_{x,y} & = & 
\Theta_{Q}(\langle z_{x,y} \rangle),\,\,\,\,
\langle z_{x,y} \rangle =  
\sum_{z}z_{x,y}{\rm Pr}(\{z\}|\{\tau\}) 
\end{eqnarray}
where $\Theta_{Q}(\cdots)$ is 
the Q-generalized step function defined by 
\begin{eqnarray}
\theta_{Q}(x) & = & 
\sum_{k=0}^{Q-1}
k 
{\Biggr \{}
\theta \left( x - \left( k - \frac{1}{2} \right) \right) - 
\theta \left( x - \left( k + \frac{1}{2} \right) \right)
{\Biggr \}}. 
\end{eqnarray}
Obviously, $\langle z_{x,y} \rangle$ 
is a local magnetization 
of the system described by 
(\ref{eq:Ham}) under 
(\ref{eq:constraint}). 
\subsection{Average case performance measure}
To investigate the performance 
of the inverse halftoning, 
we evaluate the mean square error 
which represents the pixel-wise similarity 
between the original and restored images. 
Especially, 
we evaluate the average case 
performance of the inverse halftoning 
through the following 
averaged mean square error
\begin{equation}
\sigma = 
\frac{1}{Q^2L^2}
\sum_{\{ \xi \}} {\rm Pr} \left( \{ \xi \} \right)
 \sum_{(x,y)} 
\left( \hat{z}_{x.y} - \xi_{x,y} \right)^2.
\end{equation}
We should keep in mind that 
the $\sigma$ gives zero 
if all restored images are exactly the same as 
the corresponding original images. 
\section{Results}
\label{sec:results}
In this section, 
we first investigate the statistical 
properties of our approach to the inverse halftoning 
for a set of snapshots from a Gibbs distribution of 
the ferromagnetic Q-Ising model via computer simulations.  
We next analytically evaluate 
the performance for the infinite-range model. 
Finally, we check the usefulness of our approach 
for the realistic images, namely, the 256-levels standard image 
{\it Lenna}. 
\subsection{Monte Carlo simulation}
We first carry out Monte Carlo simulations 
for a set of halftone images, which are 
obtained 
from the snapshots from a Gibbs distribution of the 
ferromagnetic $Q=4$ Ising model with $100 \times 100$ pixels 
by the uniform threshold $M_{x,y}=2\,\,
(\forall_{x,y})$ and the 
$2\times 2$ Bayer-type threshold arrays 
as shown in Fig. \ref{fig:fg2}. 
In order to clarify the statistical performance 
of our method, 
we reveal the hyper-parameters $J$ and $T_{m}$ 
dependence of the averaged mean square error $\sigma$.  
\begin{figure}[ht]
\begin{center}
\mbox{}\vspace{+0.3cm}
\includegraphics[width=9cm]{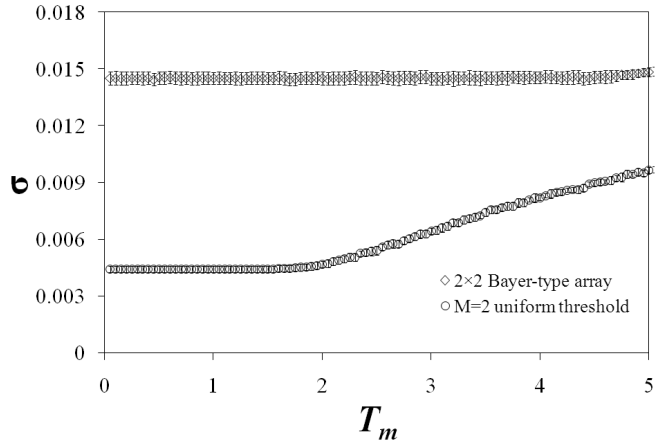}
\end{center}
\caption{\footnotesize 
The mean square error as a function of $T_m$. 
The original image is a snapshot from a Gibbs distribution of 
the $Q=4$ ferromagnetic Ising model with $100 \times 100$ pixels and 
$T_s = 1.0$, $J_s=1$ and $J=1$. 
The halftone images are obtained by the uniform 
and $2 \times 2$ Bayer-type arrays. 
}
\label{fig:fg4}
\end{figure}
We plot the results in Fig. \ref{fig:fg4}.  
These figures show that 
the present method achieves 
the best possible performance under 
the Bayes-optimal 
condition, that is, $J=J_{s}$ and $T_{m} = T_{s}$.  
We also find from Fig. \ref{fig:fg4} (the lower panel) 
that 
the limit $T_{m} \to \infty$ 
leading up to the MAP estimate 
gives almost the same performance as the Bayes-optimal 
MPM estimate. 

This fact means that 
it is not necessary for us to take 
the $T_{\rm m} \rightarrow 0$ limit 
when we carry out the inverse halftoning 
via simulated annealing. 
\begin{figure}[ht]
\begin{center}
\mbox{}\vspace{-0.3cm}
\includegraphics[width=4.5cm]{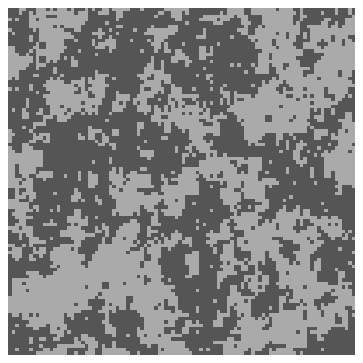}
\includegraphics[width=4.5cm]{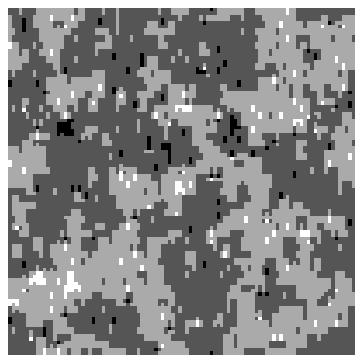}
\includegraphics[width=4.5cm]{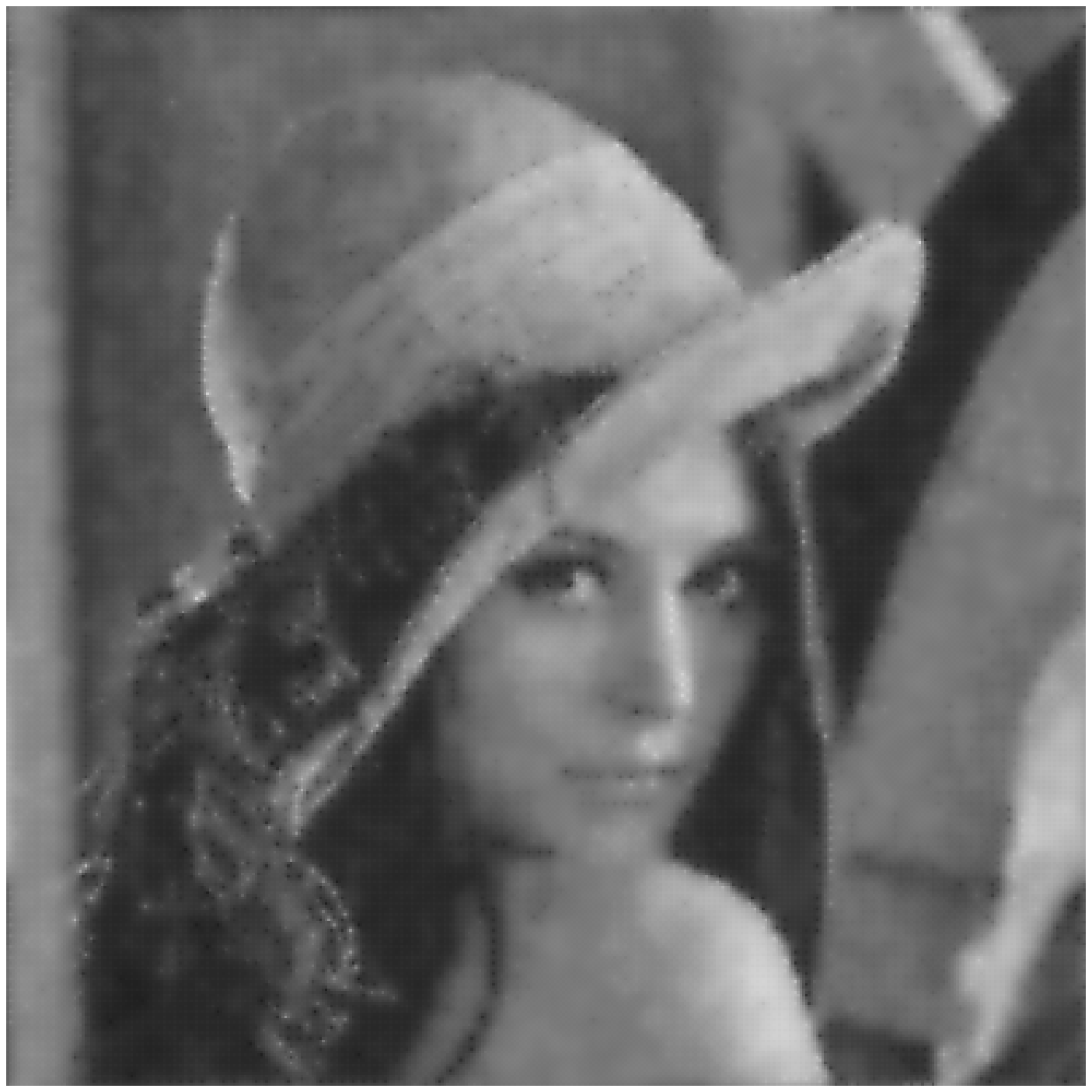}
\end{center}
\caption{\footnotesize 
The left panel shows 
a $Q=4$ grayscale image restored by the MPM estimate from the halftone 
image shown in Fig. \ref{fig:fg3} (left). 
The center panel 
shows a $Q=4$ grayscale image restored 
by the MPM estimate from the halftone 
image shown in Fig. \ref{fig:fg3} (center).  
The right panel shows  
a $Q=256$ grayscale image restored by the MPM estimate from the halftone 
image shown in Fig. \ref{fig:fg3} (right). 
}
\label{fig:fg5}
\end{figure}
\mbox{}
From the restored image in Fig. \ref{fig:fg5} (center), 
it is actually confirmed that the present method effectively works 
for the snapshot of the ferromagnetic Q-Ising model. 

It should be noted that 
the mean square error evaluated for 
the $2 \times 2$ 
Bayer-type array is 
larger than that for the $M=2$ uniform threshold. 
This result seems to be 
somewhat counter-intuitive 
because the halftone 
image shown in the center panel of 
Fig. \ref{fig:fg3} seems to be  
much closer to the original image, 
in other words, 
is much informative to retrieve 
the original image than 
the halftone image shown in 
the left panel of the same figure. 
However, 
it could be understood as follows. 
The shape of each `cluster' appearing 
in the original image 
(see the left panel of Fig. \ref{fig:fg1}) 
remains in the halftone version 
(the left panel of Fig. \ref{fig:fg3}), 
whereas, in the halftone image 
(the center panel of Fig. \ref{fig:fg3}), 
such structure is destroyed by the 
halftoning process via 
the $2 \times 2$ Bayer-type array. 
As we found, 
in a snapshot of the 
ferromagnetic Q-Ising model at 
the inverse temperature 
$J_{s}/T_{s}=1$, 
the large size clusters are much more 
dominant components than the small isolated pixels. 
Therefore, 
the averaged mean square error 
is sensitive to the change of 
the cluster size or 
the shape, and if 
we use the constant threshold mask to 
create the halftone image, 
the shape of the cluster does not 
change, whereas the high-frequency components vanish. 
These properties are desirable for us 
to suppress the increase of the averaged 
mean square error. 
This fact implies us that  
the averaged mean square error for the 
$2 \times 2$ Bayer-type 
is larger than that for the constant mask array and 
the performance is much worse than expected. 
\subsection{Analysis of the infinite-range model}
\label{subsec:InfiniteRange}
In this subsection, 
we check the validity of our Monte Carlo simulations, 
namely, we analytically evaluate 
the statistical performance of 
the present method 
for a given set of the snapshots 
from a Gibbs distribution of the ferromagnetic 
Q-Ising model in which each spin variable is located on 
the vertices of the complete graph. 
For simplicity, 
we first transform the index from 
$(x,y)$ to $i$ 
so as to satisfy $i=x+Ly+1$. 
Then, the new index $i$ runs from 
$i=1$ to $L^{2}-1=N$. 
For this new index of each 
spin variable, 
we consider the infinite-range version of true prior and 
the model as 
\begin{eqnarray}
{\rm Pr} \left( \{ \xi \} \right) & = & 
\frac{ 
{\rm e}^{ 
- \frac{\beta_{\rm s}}{2N} \sum_{i<j} ( \xi_i - \xi_j)^2}}
{Z_{\rm s}},\,\,\,
{\rm Pr} \left( \{z \} \right) =  
\frac{ 
{\rm e}^{ 
- \frac{\beta_{\rm m}}{2N} \sum_{i<j} (z_i - z_j)^2}}
{Z_{\rm m}} 
\end{eqnarray}
where the scaling factors $1/N$ appearing in front of 
the sums $\sum_{i<j}(\cdots)$ 
are needed to take a proper thermodynamic limit. 
We also set 
$\beta_{\rm s} \equiv J_{\rm s}/T_{\rm s}$ and 
$\beta_{\rm m} \equiv J/T_{\rm m}$ 
for simplicity. 
Obviously, the thermodynamics of 
the system $\{\xi\}$ is 
determined by the following magnetization: 
\begin{eqnarray}
m_{0} & \equiv & 
\frac{1}{N} \sum_{i=1}^{N}
\xi_{i} = 
\frac{
\sum_{\xi=0}^{Q-1} 
\xi
\exp [2 \beta_s m_0 \xi - \beta_s {\xi}^2 ]}
{
\sum_{\xi=0}^{Q-1} 
\exp [2 \beta_s m_0 \xi - \beta_s {\xi}^2 ]
}.
\label{eq:m0}
\end{eqnarray}
On the other hand, 
the magnetization 
for the system $\{z\}$ having 
disorders $\{\tau\}$ and $\{\xi\}$ is given 
explicitly as  
\begin{eqnarray}
m & \equiv & 
\frac{1}{N}
\sum_{i=1}^{N}
z_{i} = 
\frac{
\sum_{\xi=0}^{Q-1}
\left(
\frac{
\sum_{z=0}^{Q-1}
z\, {\rm e}^{2 \beta_m m z - \beta_m z^2}
\delta (\theta (\xi - M ), 
\theta (z - M )) 
}{
\sum_{z=0}^{Q-1}
{\rm e}^{2 \beta_m m z - \beta_m z^2}
\delta (\theta (\xi - M ),
 \theta (z - M )) 
}
\right)
{\rm e}^{2 \beta_s m_0 \xi - \beta_s {\xi}^2}}
{
\sum_{\xi=0}^{Q-1} 
{\rm e}^{2 \beta_s m_0 \xi - \beta_s {\xi}^2}
}
\label{eq:m}. \nonumber \\
\end{eqnarray}
Then, the average case performance is 
determined by 
the following averaged 
mean square error: 
\begin{eqnarray}
\sigma & \equiv & 
\frac{1}{NQ^{2}}
\sum_{i=1}^{N}
\{\xi_{i}-
\Theta_{Q}(
\langle z_{i} \rangle)
\}^{2} \nonumber \\
\mbox{} & = &  
\frac{
\sum_{\xi=0}^{Q-1}
\left\{\xi-
\Theta_{Q}\left(
\frac{
\sum_{z=0}^{Q-1}
z\,{\rm e}^{2 \beta_m m z - \beta_m z^2}
\delta (\theta ( \xi - M), 
\theta (z - M)) 
}{
\sum_{z=0}^{Q-1}
{\rm e}^{2 \beta_m m z - \beta_m z^2}
\delta (\theta ( \xi - M), 
\theta (z- M)) 
}
\right)
\right\}^{2}
{\rm e}^{2 \beta_s m_0 \xi - \beta_s {\xi}^2}}
{Q^{2} 
\sum_{\xi=0}^{Q-1} 
{\rm e}^{2 \beta_s m_0 \xi - \beta_s {\xi}^2}
} \nonumber \\
\label{eq:sigma}
\end{eqnarray}
Solving these self-consistent equations 
with respect to $m_0$ (\ref{eq:m0}) and $m$ (\ref{eq:m}),  
we evaluate 
the statistical performance of 
the present method through the quantity 
$\sigma$ (\ref{eq:sigma}) analytically. 
\par
As we have estimated using the Monte Carlo simulation, 
we estimate 
how the mean square error depends on the hyper-parameter $T_{m}$ 
for the infinite-range version of our model 
when we set to $Q=8$, $J_{s}=1$, $T_{s}=1$, $M=3.5\,(=(Q-1)/2), 4.5$ 
and $J=1$. 

\begin{figure}[ht]
\begin{center}
\mbox{}\vspace{-0.3cm}
\includegraphics[width=9cm]{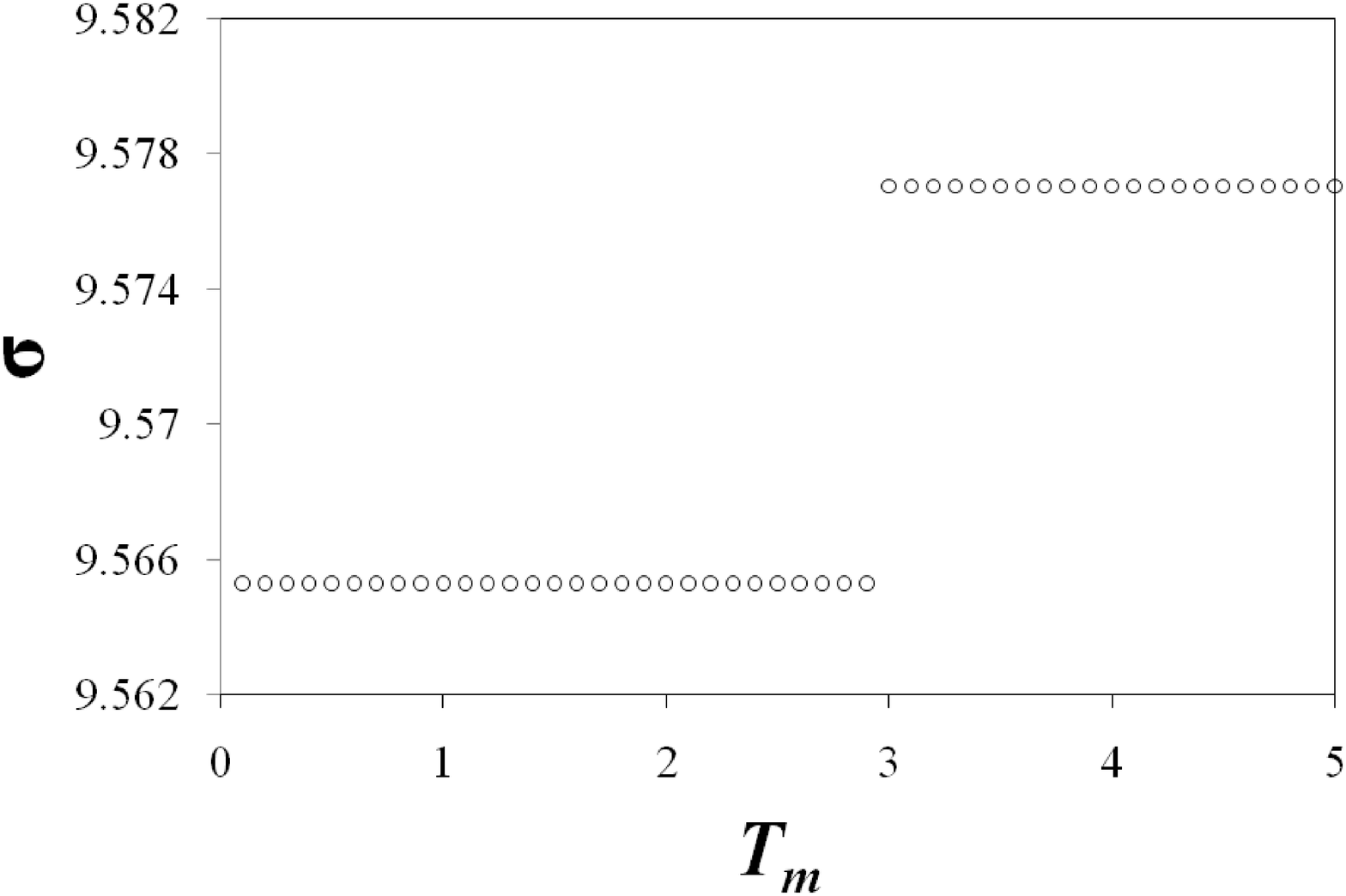}
\end{center}
\begin{center}
(a)
\end{center}
\begin{center}
\includegraphics[width=9cm]{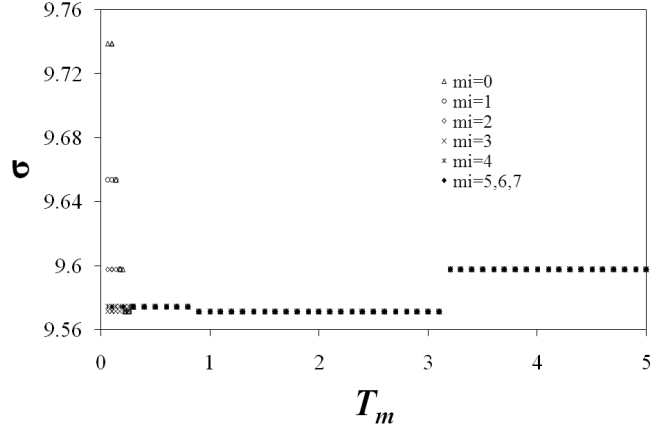}
\end{center}
\begin{center}
(b)
\end{center}
\caption{\footnotesize 
(a) The mean square error as a function of the parameter $T_m$ 
when $Q=8$, $T_s=1$, $J_s=1$, $M=(Q-1)/2$ and $J=1$, 
(b) The mean square error as a function of the parameter $T_m$ 
when $Q=8$, $T_s=1$, $J_s=1$, $M=4.5\not=(Q-1)/2$ and $J=1$. 
The value $m_{\rm i}$ for each line caption 
denotes the initial condition of 
the magnetization 
$m$ to find the locally stable solution. 
}
\label{fig:fg6}
\end{figure}

\begin{figure}[ht]
\begin{center}
\mbox{}\vspace{-0.3cm}
\includegraphics[width=9cm]{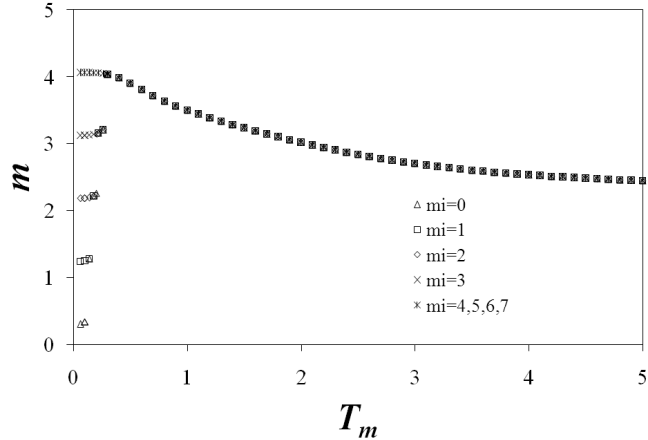}
\end{center}
\caption{\footnotesize 
The magnetization $m$ as a function of the parameter $T_m$ 
when $Q=8$, $T_s=1$, $J_s=1$, $M=4.5\not=(Q-1)/2$ and $J=1$. 
}
\label{fig:fg7}
\end{figure}

We find from Figs. \ref{fig:fg6} (a) and (b) that 
the mean square error takes its minimum 
in the wide range on $T_m$ including the Bayes-optimal condition 
$T_m = T_s \,(= 1)$. 
Here, we note that $m = m_0\, (=3.5)$ holds 
under the Bayes-optimal condition, 
$T_m = T_s$ for both cases of $M=3.5$ and $M=4.5$, 
which is shown in Fig. \ref{fig:fg7}. 
From this fact, 
we might evaluate the gap $\Delta$ between 
the lowest value of the mean 
square error and 
the second lowest value 
obtained at the higher temperature than 
$T_{s}$ as follows. 
\begin{eqnarray}
\Delta & \simeq &  
\frac{\sum_{\xi=0}^{Q-1} 
(\xi-m_{0})^{2} {\rm e}^{2\beta_{s} m_{0} \xi - 
\beta_{s} \xi^{2}}}
{Q^{2}\sum_{\xi=0}^{Q-1} 
{\rm e}^{2\beta_{s} m_{0} \xi - 
\beta_{s} \xi^{2}}}- 
\frac{\sum_{\xi=0}^{Q-1} 
(\xi-m_{0}-1)^{2} {\rm e}^{2\beta_{s} m_{0} \xi - 
\beta_{s} \xi^{2}}}
{Q^{2}\sum_{\xi=0}^{Q-1} 
{\rm e}^{2\beta_{s} m_{0} \xi - 
\beta_{s} \xi^{2}}} \nonumber \\
\mbox{} & = &  
\frac{\sum_{\xi=0}^{Q-1} 
(2\xi-2m_{0}+1){\rm e}^{2\beta_{s} m_{0} \xi - 
\beta_{s} \xi^{2}}}
{Q^{2} \sum_{\xi=0}^{Q-1} 
{\rm e}^{2\beta_{s} m_{0} \xi - 
\beta_{s} \xi^{2}}} = 
\frac{1}{Q^{2}}
\end{eqnarray}
For example, for $Q=8$, 
we evaluate the gap as $\Delta = (8)^{-2}=0.00156$ 
and this value agree with the result shown 
in Fig. \ref{fig:fg6}. 
From Figs. \ref{fig:fg6} and 
\ref{fig:fg7}, we also find that 
the range of $T_{m}$ in which 
the mean square error takes the lowest value 
coincides with 
the range of temperature 
$T_{m}$ for which 
the magnetization 
satisfies $m(T_{m})=
m(T_{s}=1) \pm 1 = 3.5\pm 1$ 
as shown in Fig. \ref{fig:fg7}. 
This robustness 
for the hyper-parameter selecting 
is one of the desirable 
properties from the view point 
of the practical use of our approach. 

Moreover, the above evaluations might be helpful for 
us to deal with the inverse halftoning 
from the halftoned image of the standard image 
with confidence.  
In fact, we are also confirmed that 
our method is practically useful 
from the resulting image 
shown in Fig. \ref{fig:fg5} (right) having the mean square error 
$\sigma = 0.002005$. 
\section{Summary}
In this paper, we investigated the condition to achieve 
the Bayes-optimal performance of inverse halftoning 
by making use of computer simulations and 
analysis of the infinite range model. 
We were also confirmed that our Bayesian approach is useful even for 
the inverse halftoning from the binary dots obtained from standard images, 
in the wide range on $T_m$ including the Bayes-optimal condition, $T_m = T_s$.
We hope that some modifications of the prior distribution might 
make the quality of the inverse halftoning much better. 
It will be our future work.  
\section*{Acknowledgment}
We were 
financially supported 
by {\it Grant-in-Aid 
Scientific Research on Priority Areas 
``Deepening and Expansion of Statistical Mechanical Informatics (DEX-SMI)" 
of The Ministry of Education, Culture, 
Sports, Science and Technology (MEXT)} 
No. 18079001. 

\end{document}